\documentclass[a4paper,11pt]{article}
\pdfoutput=1 % if your are submitting a pdflatex (i.e. if you have
\usepackage{jcappub} % for details on the use of the package, please
\usepackage[T1]{fontenc} % if needed

\usepackage[T1]{fontenc}

\usepackage[english, english]{babel}

\DeclareRobustCommand{\VAN}[3]{#2}
\let\VANthebibliography\thebibliography
\def\thebibliography{\DeclareRobustCommand{\VAN}[3]{##3}\VANthebibliography}

\usepackage{algorithm} % for algorithm environment
\usepackage[noend]{algpseudocode} % to write pseudocode

\usepackage[bitstream-charter]{mathdesign}
\usepackage[usenames]{color}
\usepackage{xcolor}
\usepackage{soul}
\usepackage{multirow}
\usepackage{graphicx}
\usepackage{wrapfig}   %
\usepackage{amsmath}
\usepackage{nicefrac}
\usepackage{microtype}
\usepackage{amssymb}
\usepackage{physics}
\usepackage{bbding}
\usepackage{csquotes} % for \enquote{}
\usepackage{float}
\usepackage{xurl}
\usepackage{adjustbox}
\usepackage{comment}
\usepackage{hyperref}
\usepackage{tensor} % for \indices{}
\usepackage{enumerate}
\usepackage{enumitem}
\usepackage{orcidlink} % ORCID
\usepackage{booktabs} % for nice tables
\usepackage[font=small,labelfont=bf]{caption}
\usepackage{subcaption}
\usepackage{pdfpages}

\hypersetup{colorlinks=true, citecolor=blue, linkcolor=blue, urlcolor=blue}

\definecolor{hotpink}{RGB}{255, 105, 180}

\definecolor{orcidlogocol}{HTML}{A6CE39}

\DeclareSymbolFont{usualmathcal}{OMS}{cmsy}{m}{n}
\DeclareSymbolFontAlphabet{\mathcal}{usualmathcal}

\newcommand{\likeli}{\mathcal{L}}

\title{\boldmath Markov Walk Exploration of Model Spaces: 
Bayesian Selection of Dark Energy Models with Supernov\ae}

\author[a,1]{Benedikt Schosser\,\orcidlink{0009-0007-8905-7749},\note{Corresponding author.}}
\author[a]{Tobias R{\"o}spel\,\orcidlink{0009-0003-8645-4643},}
\author[a,b]{and Bj{\"o}rn Malte Sch{\"a}fer\,\orcidlink{0000-0002-9453-5772}}

\affiliation[a]{Zentrum f{\"u}r Astronomie der Universit{\"a}t Heidelberg, Astronomisches Rechen-Institut,\\Philosophenweg 12, 69120 Heidelberg, Germany}
\affiliation[b]{Interdisziplin{\"a}res Zentrum f{\"u}r wissenschaftliches Rechnen, Universit{\"a}t Heidelberg,\\INF205, 69120 Heidelberg, Germany}

\emailAdd{schosser@stud.uni-heidelberg.de}

\abstract{
Central to model selection is a trade-off between performing a good fit and low model complexity: A model of higher complexity should only be favoured over a simpler model if it provides significantly better fits. In Bayesian terms, this can be achieved by considering the evidence ratio, enabling choices between two competing models. We generalise this concept by constructing Markovian random walks for exploring the entire model space. In analogy to the logarithmic likelihood ratio in parameter estimation problem, the process is governed by the logarithmic evidence ratio. We apply our methodology to selecting a polynomial for the dark energy equation of state function $w(a)$ on the basis of data for the supernova distance-redshift relation. 
}

\begin{document}
\maketitle
\flushbottom

\section{Introduction}
\label{sec:intro}
% TODO: write your article here.
In problems of statistical inference \cite{fisher_logic_1935} in cosmology \cite{trotta_bayesian_2017, trotta_bayes_2008,Hobson_Jaffe_Liddle_Mukherjee_Parkinson_2009}, Bayes'\@ theorem combines the prior information $\pi(\theta|M)$ on the parameters~$\theta$ of a physical model $M$ with the likelihood $\likeli(y|\theta,M)$ as the distribution of the data points $y$ for a given parameter choice $\theta$ to the posterior distribution $p(\theta|y,M)$,
\begin{equation}
p(\theta|y,M) = \frac{\likeli(y|\theta,M)\pi(\theta|M)}{p(y|M)}\,,
\label{eqn_bayes_theorem}
\end{equation}
with the Bayesian evidence
\begin{equation}
p(y|M) = \int\dd^n\theta\:\likeli(y|\theta,M)\pi(\theta|M)
\label{eqn_bayes_evidence}
\end{equation}
as normalisation. We denote by $M$ a member of the discrete and countable model space $\mathcal{M}$ and by $\theta$ a vector in the continuous parameter space $\Theta$ associated with $M$. In actual applications to cosmological data, use of the Fisher formalism is widespread \cite{tegmark_karhunen-loeve_1997, amara_systematic_2007, wolz_validity_2012, bassett_fisher4cast_2009, coe_fisher_2009, raveri_cosmicfish_2016}, which approximates the distributions involved in Eq.~(\ref{eqn_bayes_theorem}) as Gaussian. Analytical treatment of non-Gaussian distributions is possible by extension of the Fisher matrix formalism \cite{sellentin_breaking_2014, sellentin_fast_2015, schafer_describing_2016}. Alternatively, Markov chain Monte Carlo techniques simulate a thermal random walk on a potential given by $\ln(\likeli(y|\theta,M)\pi(\theta|M))$ with no restriction on the shape of the distributions \cite{lewis_cosmological_2002, emcee}.

\begin{figure*}[t]
    \centering
    \includegraphics[width=.75\textwidth]{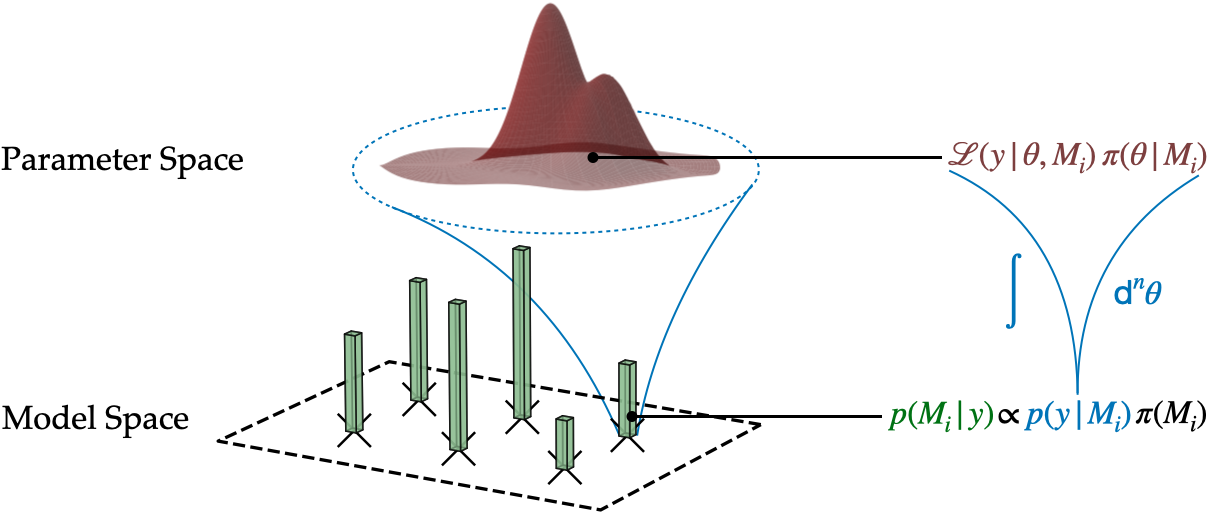}
    \caption{Conceptual overview of the two different levels of inference. For each model $M_i$ in the discrete space, a corresponding continuous parameter space exists. The discrete pdf for all models is obtained with the evidence for each model and its prior.}
    \label{fig:overview}
\end{figure*}

Model selection in the Bayesian sense requires the computation of the posterior probability $p(M|y)$ supporting a model $M$ in the light of data $y$ \cite{heavens_model_2007, kerscher_model_2019, jenkins_power_2011, trotta_applications_2007, trotta_forecasting_2007,Hee_2015}. Application of Bayes' theorem allows the inversion of the conditional probability $p(y|M)$ to give $p(M|y)$, according to
\begin{equation}
p(M|y) = \frac{p(y|M)\pi(M)}{p(y)}\,.
\end{equation}
The normalisation is provided by the model evidence $p(y)$ and would result from summation over the discrete space of models $M_i$, each with an associated model prior probability $\pi(M_i)$ and a Bayesian evidence $p(y|M_i)$ according to Eq.~(\ref{eqn_bayes_evidence}).
\begin{equation}
p(y) = \sum_i p(y|M_i)\pi(M_i)\,.
\label{eqn_model_evidence}
\end{equation}
Typically, two or more models are compared by taking their evidence ratios \cite{liddle2006model}. The Bayes' ratio ${B = \ln[ p(y|M_i) /  p(y|M_j)]}$ is then interpreted following Jeffreys' guideline \cite{Jeffreys:1939xee}. Other interpretations, such as promoting it to a frequentist statistic, are also explored in the literature \cite[compare][as an application]{amendola2024distributionbayesratio, desicollaboration2024desi}. To get the properly normalised posterior, one would have to calculate the evidence for every model and then compare. For large model spaces and typical cosmological problems with many parameters, this seems unfeasible. The motivation of our paper is the evaluation of the posterior probabilities $p(M_i|y)$ by generalisation of a pairwise model comparison based on Bayesian evidences. This effectively requires a Markovian process that explores the model space and is steered by the logarithmic evidence ratio. In the spirit of a Metropolis-Hastings-like algorithm, one would want to establish an ergodic random walk across the discrete landscape of models $M_i$, with a probability for transitioning from model $M_i$ to model $M_j$ proportional to the Bayesian evidence ratio $p(M_{i}|y)/p(M_{j}|y)$. The two layer structure and general idea is displayed in Fig.~\ref{fig:overview}. The full model posterior could then be used to perform Bayesian model averaging, to include model uncertainty in parameter inference \cite{10.1093/mnras/stae101, paradiso2024evaluatingextensionslcdmapplication}. Contrary to symbolic regression (SR) \cite{koza1992genetic, Udrescu:2019mnk, Cranmer:2020wew} we use a clear distinction between models and parameters, i.e. a different parameter value does not correspond to a new model. In analogy to canonical partitions $Z[\beta,J|M]$ for the exploration of parameter spaces and the computation of posterior probability distributions $p(\theta|y,M)$ for model parameters $\theta$ within a fixed model $M$, we aim to derive analogous partitions $Y[\zeta]$ for posterior model probabilities $p(M_i|y)$, as an analytical description of the sampling process for the evaluation of $p(M_i|y)$.

Our paper is structured as follows: After summarising the principles of Bayesian model selection in Sect.~\ref{sec:intro}, we implement a Markov sampler for exploring model spaces in Sect.~\ref{sec:markov}. We discuss prior choices and their relation to information criteria in Sect.~\ref{sec:prior_choice}. A theoretical excursion is made in Sect.~\ref{sec:bayes_model} by working out the theory of partition functions for model spaces. We try out our methodology in Sect.~\ref{sec:toy_model} to a toy example and apply it to the problem of selecting a polynomial model for the dark energy equation of state as a topical example from cosmology in Sect.~\ref{sec:supernova} \cite{Chung2003, tsujikawa_quintessence:_2013, liddle_present_2006, Dhawan_2017}. We discuss our main results in Sect.~\ref{sec:conclusion}.

Throughout the paper, we adopt the summation convention and denote parameter tuples $\theta^\alpha$ and data tuples $y^i$ as vectors with contravariant indices; Greek indices are reserved for quantities in parameter space and Latin indices for objects in data space. Concerning the cosmological example, we assume a flat, dark energy-dominated Friedmann-Lema{\^i}tre-Robertson-Walker spacetime with a dark energy equation of state parameter $w$, which we assume to be polynomial in the scale factor $a$.

%%%%%%%%%%%%%%%%%%%%%%%%%%%%%%%%
%%%%%%%%%%%%%%%%%%%%%%%%%%%%%%
\section{Markov exploration of model spaces}
\label{sec:markov}
% ---  --- %
\subsection{Evaluation of model posterior probabilities}
A pathway of computing the posterior model probability $p(M|y)$ from the Bayesian evidence $p(y|M)$ would be a Markov process in the discrete space of models $M_i$. In this model landscape, every model $M_i$ is located at a site $i$ and has an associated Bayesian evidence $p(y|M_i)$, commonly computed through nested sampling \cite{Buchner_2023, Ashton_2022, Skilling_2006, Feroz_2009}. The model landscape is effectively a graph that connects similar, related models to each other. The edges of this graph would represent the changes to a model when transitioning from one site $i$, or node, to another site $j$. Physically, this would correspond to adding a certain degree of freedom to the model, or removing a degree of freedom from the model. In short, the topology of the model space, as defined by the connectivity of the graph, orders the model space by complexity and neighbouring models are physically related to each other, as they possess similar features.

Even though establishing detailed balance for a walk on this graph is not straightforward, there is an advantage of ordering models by complexity: Commonly, these are penalised by their prior probability $\pi(M_i)$ and should not be favoured by model selection unless they are providing good fits to data. Adding further complication to a badly fitting already complex model would not be sensible, and these paths should be avoided in exploring the model space.

If the probability for transitioning from site $i$ to site $j$ is determined by the evidence ratio reminiscent of the conventional likelihood ratio of the Metropolis-Hastings algorithm \cite{metropolis_monte-carlo:_1985, Hastings_1970}, the number of visits of a site $i$ in this sampling process is then proportional to the posterior probability $p(M_i|y)$. In this way, the pairwise comparison of two models in the Bayesian evidence ratio is generalised to the computation of the posterior probability. From the numerics point of view, the normalisation $p(y)$ needed in $p(M_i|y) = p(y|M_i)\pi(M_i)/p(y)$ is replaced by the total number of samples, and it would be unnecessary to carry out the summation over the entire space of models, as only realistic models are taken into consideration in the motion of the Markov chain: Non-viable models with a low Bayesian evidence would be visited by the Markov walk at low probability or not at all. In this Bayesian inversion, we effectively replace the probability ratio, or "odds" of a pairwise model comparison to a properly normalised posterior probability for the entire model space.

A Metropolis-Hastings algorithm would determine the transition from model $i$ to model $j$ by means of the criterion,
\begin{equation}
p_{i \rightarrow j} = 
\mathrm{min}\left[1,  \frac{p(y|M_j) \pi(M_j)}{p(y|M_i) \pi(M_i)} \right]\,.
\end{equation}
Finding the correct stationary distribution of a Markov chain Monte Carlo (MCMC) requires two conditions, $(i)$ ergodicity and $(ii)$ detailed balance. Ergodicity means that every state is reachable in a finite number of steps. Here, this is fulfilled since the transition probabilities are non-zero, albeit they might be very small. Detailed balance requires a more careful consideration on this non-trivially shaped discrete space. For a symmetric proposal distribution $g(M_i|M_j)$ it is automatically fulfilled. However, if $g$ is not symmetric one needs a correction factor for the acceptance probability $A$. It gives the probability to jump from model $M_i$ to $M_j$ as
\begin{equation}
    A(M_i, M_j) = \mathrm{min}\left[1, q_{i \rightarrow j} \right]
    \label{eq:acceptance_probability}
\end{equation}
with
\begin{equation}
q_{i \rightarrow j} =\frac{p(y|M_j) \pi(M_j)}{p(y|M_i) \pi(M_i)} \frac{g(M_i|M_j)}{g(M_j|M_i)} \,.
\end{equation}
The ratio of the proposal distributions counteracts the situation that some states have more connections to others, making it less likely to reach them. Here, we want to point out similarities and differences between our method and Green's reversible jump MCMC (RJMCMC) \cite{Green_RJMCMC}. Our goal is to obtain $p(M|y)$ while RJMCMC samples from $p(M,\theta|y)$, which does not use the two layer structure of model and parameter space, as in every step one can change the model. Averaging over $\theta$ would build a connection between the two methods. Additionally, we do not use the assumption of nested models and obtain detailed balance without derivatives but with combinatorial corrections. At this point we would like to emphasise that experiments can be designed with optimised differentiating ability between competing models \cite{amara_model_2014}. Bayesian model selection is also performed in the field of SR \cite{Bartlett:2023gvh}, but there are differences between the approaches. We explicitly do not need to perform an exhaustive model search, but limit ourselves to the shape of polynomial models (see Sect.~\ref{sec:model_space}). This limitation can be reduced in future work. 

%  ---  ---%
\subsection{Construction of a model space}
\label{sec:model_space}
\begin{wrapfigure}{r}{0.5\textwidth}
    \centering
    \includegraphics[width=0.48\textwidth]{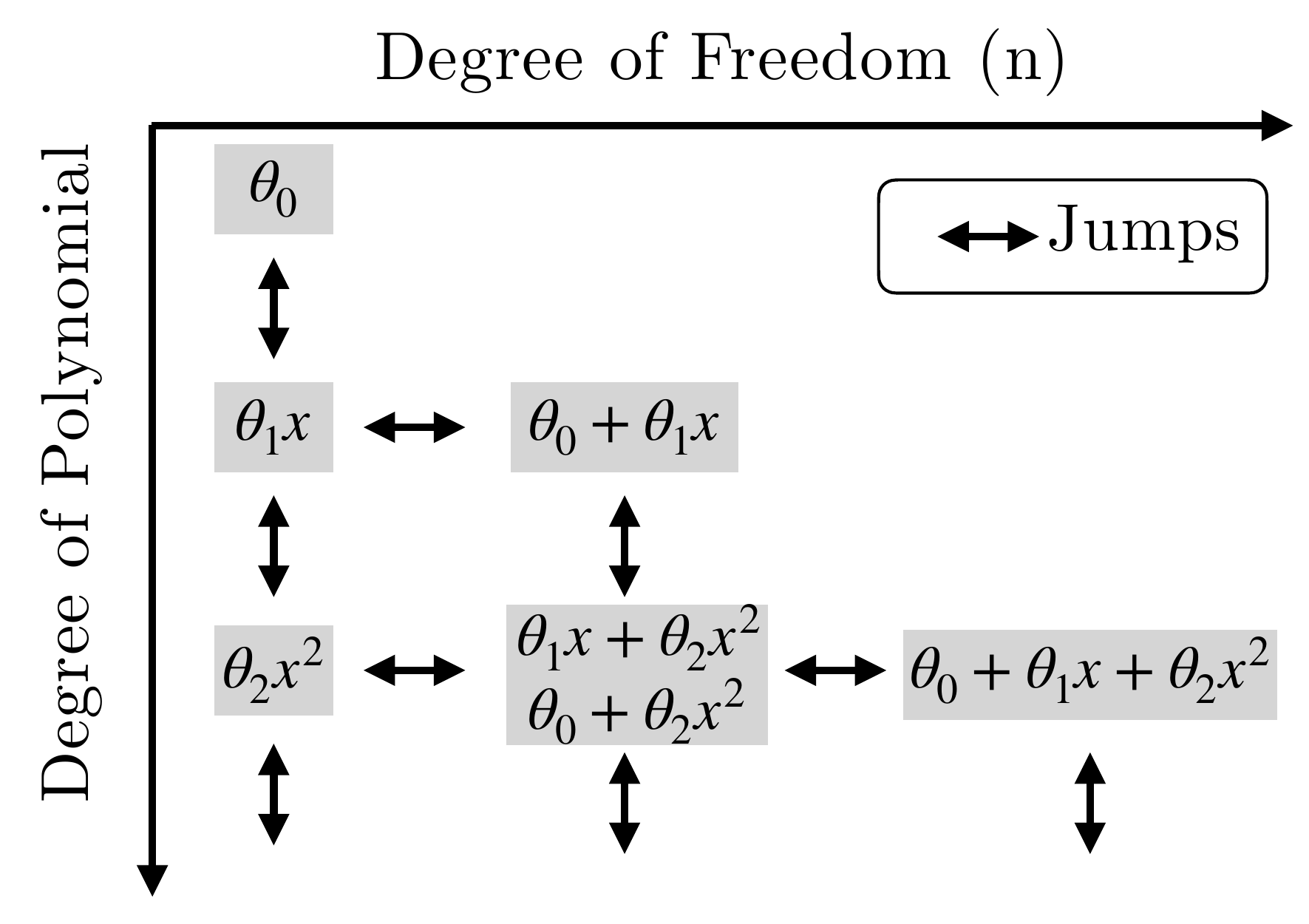}
    \caption{Outline of the discrete two-dimensional model space of polynomials.}
    \label{fig:model_space_polynoms}
\end{wrapfigure}
For our particular application in cosmology we choose the dark energy equation of state function $w(a)$ to be a polynomial in scale factor $a$. However, this model space is applicable for all polynomial models, as we illustrate in our toy example in Sect.~\ref{sec:toy_model}. To navigate this discrete space of polynomials we need a layout to order the polynomials according to some scheme. We focus on the number of free parameters and the highest polynomial degree, illustrated in Fig.~\ref{fig:model_space_polynoms}. It closely resembles Pascal's triangle, where polynomials with equal properties are at the same position and their abundance is given by the respective entry in Pascal's triangle. We allow movement in the space by either changing the number of degrees of freedom (defined as the number of monomials), or the highest degree of the polynomial, which should give a resemblance between neighbouring sites in the model space and an intuitive similarity between them.

For one proposal step, we allow only one of those to change, which is picked at random. If it is decreased or increased is also chosen randomly. If one of those is not possible, because e.g. the model has only one parameter, the other one is chosen. The more complex the models become the more models have the same number of d.o.f. and the highest degree. For example, there are three polynomials with two parameters and the highest degree is three (see Fig.~\ref{fig:model_space_polynoms}). When a step is proposed to such a state, the actual model is drawn randomly from this sample. The number of models at one state is given by the binomial coefficient $\binom{d}{\tilde{n}}$, where $d$ is the row and $\tilde{n}$ is the column of the triangle. Their smallest value is 0 and we call it $\tilde{n}$, because it is equal the degree of freedom minus one of the models.
Later $n=\tilde{n}+1$ is the d.o.f. of each model. The layout of the model space groups similiar models close to each other, yet, it can happen that the data is poorly explained by all neighbouring models. The chain only explores the local minimum. To overcome this issue, we allow multiple steps at once before evaluating the model. The number of steps is sampled from a Poissonian.

The boundaries and the degeneracy yield the asymmetry of the proposal distribution $g(M_i|M_j)\neq g(M_j|M_i)$. To derive the correct balancing factor in Eq.~(\ref{eq:acceptance_probability}) we introduce a rescaled proposal function 
\begin{equation}
\tilde{g}(j|i) = 
c_{i\rightarrow j} g(j|i) = \mathrm{const.} = c_{j\rightarrow i} g(i|j) = \tilde{g}(i|j) 
\quad \forall i,j \,,
\end{equation}
where $i$ and $j$ indicate the initial and final model, respectively. Here, $g(j|i) = g(M_j|M_i)$ is the proposal probability to jump from $M_i$ to $M_j$, $c_{i\rightarrow j}$ is the correction prefactor which accounts for the geometry of the model space, it depends on the number of borders and the degeneracy at each state. The prefactors $c_{i\rightarrow j}$ depend on these states and have to be calculated. To calculate the rescaling factor we can start by looking at one path going from an initial state $i$ to a final state $j$ via intermediate states $m$. We need to consider two corrections, the boundaries and the degeneracy. At a boundary site, one of $d$ or $n$ cannot decrease or increase, which reduces the number of possible moves by two. To preserve detailed balance we include a factor of $1/2$ to compensate for the missing possibilty. The number of models with the same $d$ and $\tilde{n}$ is given by the binomial coefficient, as the positions are closely related to Pascal's triangle. The considered model is randomly chosen at the position in the triangle, yielding the probability of one over the binomial coefficient. To counteract this we need to include the binomial coefficient for every state on the path, yielding the following formulas
\begin{align}
    &\left. c_{i \rightarrow j} \right|_\mathcal{P}=  c_{i \rightarrow M_1}  \cdot \cdot \cdot c_{M_N \rightarrow j} = \left( \frac{1}{2}\right)^{b_i} \binom{d_{M_1}}{\tilde{n}_{M_1}} \cdot \cdot \cdot \binom{d_{M_N}}{\tilde{n}_{M_N}} \binom{d_{j}}{\tilde{n}_j} \nonumber \\
    &\left. c_{j \rightarrow i} \right|_\mathcal{P}=  c_{j \rightarrow M_N}  \cdot \cdot \cdot c_{M_1 \rightarrow i} = \left( \frac{1}{2}\right)^{b_j} \binom{d_{M_N}}{\tilde{n}_{M_N}} \cdot \cdot \cdot \binom{d_{M_1}}{\tilde{n}_{M_1}} \binom{d_{i}}{\tilde{n}_i}\,, 
    \label{eq:correction_factors}
\end{align}
where $\tilde{n}$ and $d$ are the respective positions of the state on the triangle. $N$ is the number of steps taken. $b_i$ and $b_j$ are the number of borders of the initial and final state, respectively. The upper top of the triangle is a special case, where we count three borders. Eq.~\ref{eq:correction_factors} is obtained by considering one examplary path of $N$ random steps and considering the possible moves at each intermediate model. At each site one has to count the degeneracies, giving rise to the binomial factors, and the borders at start and end point, yielding the $1/2$ prefactors. Getting the correct factor demands summing over all possible paths and then taking their ratio. However, we can see that the ratio does not depend on the path, as all the binomial coefficients with $m$ as subscripts cancel. Therefore, the final relation is
\begin{equation}
\frac{g(j|i)}{g(i|j)} = \left(\frac{1}{2}\right)^{b_j-b_i}\frac{\binom{d_i}{\tilde{n}_i}}{\binom{d_j}{\tilde{n}_j}}\,.
\end{equation}

%  ---  ---%
\subsection{Binary keys for addressing the model space}\label{subsec:bin_keys}
Algorithmically, we set up a binary key $\kappa$ for addressing the model space: The length of the key, i.e. the total number of digits $d$ indicates the highest power of the polynomial, and the individual digits determine whether a certain monomial is part of the model. For instance, $\kappa = (0,1,1)$ would be the polynomial with a linear term and a quadratic term, but be missing a constant term, as indicated by the values in the tuple $\kappa$. The number of digits equal to one corresponds to the model dimensionality $n$, as they count the number of coefficients $\theta^\mu$ that are available for fitting the model to data. The possible ways of moving in model space correspond to $(i)$ changing $d$ by one or $(ii)$ changing $n$ by one.

Closely related to the binary key is the expectation value of individual key bits $\kappa^\alpha$. In the case of a model space consisting of polynomials, these probabilities would contain the information whether a certain monomial should be included in the model. With the summation running over the space of keys instead of the physical models, the expectation value is straightforwardly given by
\begin{equation}
    \bra \kappa^\alpha\ket = \sum_\mathrm{keyspace} p(\kappa^\alpha|y) \kappa^\alpha \,.
    \label{eq:mean_keyspace}
\end{equation}

In Algorithm \ref{alg:algorithm} we provide a summary of the working principle of our sampling algorithm in pseudocode. The Python implementation for both the polynomial model landscape and an actual application to cosmology data is available on \href{https://github.com/cosmostatistics/ebms_mcmc}{GitHub}.
\begin{algorithm}[t]
\caption{Markov Walk Exploration of Model Spaces}
\label{alg:algorithm}
\begin{algorithmic}[1]
    \Require Initial model $M^{(0)}$, data $y$, model prior $\pi(M)$, parameter priors $\pi(\theta \mid M)$, Poisson rate $\lambda$, number of iterations $N$
    \Ensure Empirical estimate of $p(M \mid y)$

    \State Initialize storage for evidence estimates $\{\hat{p}(y \mid M)\}$

    \For{$t = 1$ to $N$}
        \State $M_{\text{curr}} \gets M^{(t-1)}$
        \State Draw $K \sim \mathrm{Poisson}(\lambda)$
        \State $M_{\text{prop}} \gets M_{\text{curr}}$
        
        \For{$k = 1$ to $K$}
            \State Propose a single-edge move $M_{\text{prop}} \to M'$ among its neighbors
            \State $M_{\text{prop}} \gets M'$
        \EndFor
        
        \If{$\hat{p}(y \mid M_{\text{prop}})$ not computed}
            \State Compute $\hat{p}(y \mid M_{\text{prop}})$ (e.g.\ via nested sampling)
        \EndIf
        
        \State Compute Hastings ratio:
        \[
        r = 
        \frac{\hat{p}(y \mid M_{\text{prop}}) \, \pi(M_{\text{prop}})}
             {\hat{p}(y \mid M_{\text{curr}}) \, \pi(M_{\text{curr}})}
        \times
        \frac{g(M_{\text{curr}} \mid M_{\text{prop}})}
             {g(M_{\text{prop}} \mid M_{\text{curr}})}
        \]
        
        \State $A \gets \min\{1, r\}$
        \State Draw $u \sim \mathrm{Uniform}(0,1)$
        
        \If{$u < A$}
            \State $M^{(t)} \gets M_{\text{prop}}$
        \Else
            \State $M^{(t)} \gets M_{\text{curr}}$
        \EndIf
    \EndFor
    
    \State \Return Empirical frequencies of $\{M^{(t)}\}$ as estimate of $p(M \mid y)$
\end{algorithmic}
\end{algorithm}

% ---  --- % 
\section{Prior choices}\label{sec:prior_choice}
Going through the full inference pipeline from data to the model posterior requires two different priors, one for the parameters of each model $\pi(\theta|M_i)$ and one for the model itself $\pi(M_i)$. As is the case in every Bayesian analysis, one needs to make a decision about which prior to choose and motivate why it is sensible, even though in the choice of the specific functional form one can be guided by maximum entropy considerations \cite{handley_maximum_2018}.

The model prior could be chosen to reflect model complexity or the number of parameters of $M_i$,
\begin{equation}
\label{eq:one_over}
\pi_{\text{OVN}}(M_i) = \frac{1}{n}\,.
\end{equation}
In a more elaborate way, one can relate the model maximum a posteriori (MAP) $\hat{M}$ in zero order of the parameter likelihood to information criteria such as the Bayes or Akaike information criterion \cite{akaike_ic,bayesIC} as
\begin{equation}
\begin{split}
    \hat{M} &= \underset{M}{\mathrm{argmax}} \left(\ln p(M|y) \right) \\ &= \underset{M}{\mathrm{argmax}} \left( \ln p(y|M) + \ln \pi(M) - \ln p(y) \right)\,.
\end{split}
\end{equation}
Ignoring the evidence $p(y)$ as it is the same for all models, plugging in the model evidence $p(y|M)$ and multiplying by an arbitrary factor of $-2$ because the information criteria aim to minimise, one gets
\begin{equation}
\begin{split}
    \hat{M} &= \underset{M}{\mathrm{argmin}} \left(-2 \ln \int \dd^n \theta \likeli (y | \theta, M)  \pi (\theta | M) - 2 \ln \pi (M) \right)\,, 
\end{split}
\end{equation}
where $\pi (\theta | M)$ denotes the prior for the parameters of each model. Approximating the likelihood $\likeli (y | \theta, M) $ in zero order by the corresponding maximum likelihood estimate $\hat{\likeli}(y| M) = \likeli (y | \hat{\theta}, M)$, one can rewrite the equation as
\begin{equation}
\begin{split}
    \hat{M} &= \underset{M}{\mathrm{argmin}} \left(-2 \ln \hat{\likeli} (y| M) \underbrace{\int \dd^n \theta \pi (\theta | M)}_{=1} - 2 \ln \pi (M) \right) \\ &= \underset{M}{\mathrm{argmin}} \left(-2 \ln \hat{\likeli} (y| M) - 2 \ln \pi (M) \right)\,.
\end{split}
\end{equation}
This motivates for the information criteria (IC) to hold
\begin{equation}
    IC = -2 \ln \hat{\likeli} (y| M) - 2 \ln \pi_{IC} (M) \,.
\end{equation}
This flat-likelihood approximation is very strong and its purpose is purely to make a connection to the well-established information criteria. It is not used in the calculation of evidences. The normalisation of the prior is not important, however, the normalisability is. It reads 
\begin{equation}
    N_P = \sum_{d=0}^\infty \sum_{n=0}^d \binom{d}{n} \pi(M_i)\,,
\end{equation}
which diverges for all of the priors derived from information criteria and listed in Tab.~\ref{tab:IC_priors}. Therefore, we need to introduce a $d_\text{max}$, the maximal degree of the polynomial, setting an upper limit. Finally, we consider one prior, which is non-zero for every model and normalisable. It not only punishes many parameters but also a high degree. This coincides with the intuitive approach of favouring simpler models. It reads
\begin{equation}
    \pi_\text{NP}(d,n) = \frac{1}{(d+1)^{n+1}}\,,
    \label{eq:norm_model_prior}
\end{equation}
where convergence of the series can be shown easily. This specific choice for the prior is motivated by objective to have a normalisable one without sharp boundaries as well as the intuitive punishing arguments. Eq.~\ref{eq:norm_model_prior} incorporates this motivation as it decays as slow as possible while still converging and keeping the intuition. 

Tab.~\ref{tab:IC_priors} shows the results for the Akaike information criterion (AIC), the Bayesian information criterion (BIC) as well as in the inverse direction for the model prior defined in Eq.~(\ref{eq:one_over}) (OVN - one over $n$). Additionally, the normalisable prior (NP) as well as a uniform prior (U) are considered.
\begin{table}[H]
    \centering
    \begin{tabular}{@{}lllllll@{}}
        \toprule
               & AIC & BIC & OVN & NP & U\\ \midrule
        IC     & $\hat{\chi}^2 + 2n$ & $\hat{\chi}^2 + 2n \ln N$ & $\hat{\chi}^2 + 2 \ln n$ & $\hat{\chi}^2+ (n+1) \ln(d+1)$ & $\hat{\chi}^2$\\
        $\pi$  & $\mathrm{exp}(-n)$ & $N^{-n/2}$ & $1/n$ & $1/(d+1)^{n+1}$  & $1$\\ \bottomrule
    \end{tabular}
    \caption{Information criteria (IC) and their corresponding priors ($\pi$) with $n$ being the number of parameters, $N$ the number of data points and $d$ the degree of the polynomial. It is important to note that the model priors are not normalised.}
    \label{tab:IC_priors}
\end{table}
For the prior on the parameters of the model $\pi(\theta|M_i)$ information from previous, independent experiments can be used, as well as considerations from theory. In this work, we consider two different prior distributions, uniform and Gaussian. We choose those, because they are both entropy-maximising distributions under certain conditions and wide prior parameter ranges should not have a strong biasing effect on the posterior distribution. In addition, they avoid complications along the reasoning of the Jeffreys-Lindley paradox. But even in cases where the limits of the prior do not affect the resulting posterior, they still can affect the value of the evidence through volume effects. Therefore, one cannot make the prior as wide as numerically possible without good reason but has to carefully consider the effect on the Bayesian evidence.

% ---  --- %
\section{Partition functions for Bayesian model selection}
\label{sec:bayes_model}
Before applying the model selection algorithm, we want to make a theoretical remark on an analogy to partition functions originating from statistical physics, which could allow to fine-tune the algorithm via an introduced temperate $\zeta$ or to derive interesting quantities by taking derivatives.
Having a look at parameter inference, the structure of Bayes'\@ theorem in Eq.~\ref{eqn_bayes_theorem} with an integral (cf. Eq.~\ref{eqn_bayes_evidence}) in the denominator and the integrand in the numerator suggests the definition of the canonical partition function \cite{jaynes_information_1957}
\begin{equation} \label{eq:canZ}
Z[\beta,J|M] =
\int\dd^n\theta\:\left[\likeli(y|\theta,M)\pi(\theta|M)\:\exp(J_\alpha\theta^\alpha)\right]^\beta\,,
\end{equation}
which falls back on the Bayesian evidence $p(y|M)$ for $\beta = 1$ and $J = 0$. By differentiation of the Helmholtz free energy
\begin{equation}
F(\beta,J|M) = -\frac{1}{\beta}\ln Z[\beta,J|M]
\label{eqn_helmholtz}
\end{equation}
with respect to $J_\alpha$, cumulants of the posterior distribution $p(\theta|y,M)$ can be computed, making them easily accessible beyond second order \cite{10.1093/mnras/Rover, kuntz2023partition, partitionfunction101}.
% ---  --- %
\subsection{Canonical partitions on model spaces}
In complete analogy to the canonical partition $Z[\beta,J|M]$ as an analytical description of an MCMC sampling process in the parameter space of the model $M$, one can introduce a second, likewise canonical partition $Y[\zeta]$ for a sampling process in the abstract, discrete space of models $M_i$,
\begin{equation}
Y[\zeta] = \sum_i [p(y|M_i)\pi(M_i)]^\zeta\,,
\label{eqn_partition_y}
\end{equation}
as a generalisation to Eq.~(\ref{eqn_model_evidence}). The partition function $Y[\zeta]$ falls back on the model evidence $p(y)$ for the choice $\zeta = 1$, as the proper normalisation of the posterior model probability $p(M|y) = p(y|M)\pi(M)/Y[\zeta = 1]$. For generality, we introduce a second temperature $\zeta$ that controls the sampling process in the model space, as there might be reasons to choose it differently from the temperature $\beta$ entering $Z[\beta,J|M]$.

Potentially different temperatures $\beta$ and $\zeta$ illustrate the fact that the partition $Y[\zeta]$ relevant for Bayesian model selection is actually a partition of partitions, as it generalises to:
\begin{align}
Y[\zeta,\beta] = & 
\sum_i \left(\int\dd^n\theta\left[\mathcal{L}(y|\theta,M_i)\pi(\theta|M_i)\right]^\beta\pi(M_i)\right)^\zeta\nonumber\\
=& \sum_i\left(Z[\beta|M_i]\pi(M_i)\right)^\zeta\,,
\end{align}
where $Z[\beta|M]$ is evaluated at $J=0$. The two levels of the partition function are visualised in Fig.~\ref{fig:overview}. Low values of $\zeta$ (or $\beta$) would enable the Markov process to explore the model space (or parameter space) more extensively, but only the value $\zeta = 1 = \beta$ is compatible with Bayes' law. Slow increase of $\zeta$ towards unity would effectively be a simulated annealing process, and arguably annealing in the model space would only be sensible in varying $\zeta$ while keeping $\beta=1$ fixed.

\subsection{Entropy, specific heat and surprise statistic}
Not only can the Markov process be shaped by the partition function approach, but it also allows for an efficient computation of interesting quantities like the entropy, specific heat or surprise statistics. Following the definition of the Helmholtz free energy $F(\beta,J|M)$ leads to an associated thermodynamic potential $\Phi(\zeta)$
\begin{equation}
\Phi(\zeta) = -\frac{1}{\zeta}\ln Y[\zeta]\,.
\end{equation}
As expected for canonical partitions, the entropy $S$ can be isolated through
\begin{equation}
S(\zeta) = -\frac{1}{\zeta^2}\frac{\partial\Phi(\zeta)}{\partial\zeta}
\end{equation}
and corresponds exactly to Shannon's information entropy of the model posterior probability $p(M_i|y)$,
\begin{equation}
S = - \sum_i p(M_i|y) \ln p(M_i|y) = \left\langle\ln p(M_i|y)\right\rangle
\quad \text{at}\quad \zeta = 1\,.
\end{equation}
The second derivative of $\Phi(\zeta)$ with respect to temperature, which corresponds in thermodynamical applications to the specific heat $C$, becomes
\begin{equation}
C = 
-\zeta\frac{\partial S}{\partial\zeta} = 
\left\langle \ln^2p(M_i|y)\right\rangle - \left\langle\ln p(M_i|y)\right\rangle^2
\quad \text{at}\quad \zeta = 1\,,
\end{equation}
which is strictly positive and equal to the variance of the logarithmic posterior.

Constructing a related partition
\begin{equation}
Y^\prime[\zeta] = \sum_i p(y|M_i)^\zeta \pi(M_i)^{(\zeta+1/\zeta)/2}
\end{equation}
would lead through differentiation of $\ln Y^\prime/\zeta$ with respect to $\zeta$ to 
\begin{equation}
\Delta S = \sum_i p(M_i|y)\ln\frac{p(M_i|y)}{\pi(M_i)}
\quad\text{at}\quad \zeta = 1\,,
\end{equation}
recovering the surprise statistic applied to model prior and posterior \cite{Kuntz:2024vdd}. It is defined as the Kullback-Leibler divergence between the prior and posterior. As such, it would reflect the information gain in transitioning from the prior to the posterior distribution, applied to the space of all models \cite{lindley_information_1956}.

\subsection{Model partition sum for binary key}
The partition sum formalism becomes particularly instructive when applied to the binary key introduced in Sect.~\ref{subsec:bin_keys}. In the case of estimating model parameters $\theta^\mu$ for a given model, one could use the canonical partition $Z[\beta,J|M]$ (Eq.~\ref{eq:canZ}) and the associated Helmholtz energy $F(\beta,J|M)$ (Eq.~\ref{eqn_helmholtz}) to generate cumulants of the posterior distribution $p(\theta|y,M)$ by differentiation of $F(\beta,J|M)$ with respect to $J_\alpha$. In an analogous way, the binary key $\kappa$ enables an extension of the partition $Y[\zeta]$ in Eq.~(\ref{eqn_partition_y}) by a "model source" $K$ to
\begin{equation}
Y[\zeta,K] = \sum_\text{keyspace} [p(y|\kappa)\pi(\kappa)]^\zeta \exp(\zeta K_\alpha \kappa^\alpha)\,,
\end{equation}
if the models are not simply numbered by $i$, but addressed by their binary key. Again, the summation runs over the space of keys instead of the physical models. Then, differentiation of $\ln Y[\zeta,K]/\zeta$ with respect to $K_\alpha$ gives the expectation value of individual key bits $\kappa^\alpha$,
\begin{equation}
\frac{\partial}{\partial K_\alpha}\frac{\ln Y [\zeta,K]}{\zeta} = 
\frac{\sum_\mathrm{keyspace} p(\kappa^\alpha|y) \kappa^\alpha}{\sum_\mathrm{keyspace} p(\kappa|y)} \bigg|_{K=0, \zeta=1} = 
\bra \kappa^\alpha\ket\,,.
\label{eq:ps_mean_keyspace}
\end{equation}
This alternative approach to obtain the information whether to include a monomial in the model concludes the excursion to model partition sums. It can be seen as a theoretical framework for the numerical approach to the underlying task. All of the following results are obtained with the algorithm described in Alg.~\ref{alg:algorithm}. The thermodynamic description may be used for further theoretical investigations.

% ---  --- %
\section{Linear toy model}\label{sec:toy_model}
It is instructive to test our method on a toy model with known ground truth. Therefore, polynomial data with different degrees, number of parameters and noise levels are generated. With respect to the parameters $\theta$ the polynomial toy model is linear and thus can be written without the added noise as
\begin{equation}
    y_\mathrm{Model}^i = A\indices{^i_{\alpha}} \theta^{\alpha}.
\end{equation}
This results into the $\chi^2$ given by
\begin{equation}
    \begin{split}
        \chi^2 &= (y^i - A\indices{^i_{\alpha}} \theta^{\alpha}) C_{ij} (y^j - A\indices{^j_{\beta}} \theta^{\beta}) \\
        &= y^i C_{ij} y^j - 2 \underbrace{y^i C_{ij} A\indices{^j_{\alpha}}}_{:= Q_{\alpha}}  \theta^{\alpha} + \underbrace{A\indices{^i_{\alpha}} C_{ij} A\indices{^j_{\beta}}}_{:= F_{\alpha \beta}} \theta^{\alpha} \theta^{\beta}\,,
    \end{split}
\end{equation}
where $F_{\alpha \beta}$ is the Fisher matrix \cite{rover2023partition}. The data likelihood is actually Gaussian because Gaussian noise is used in the creation of our test samples. This allows to test for different priors, but Gaussian priors in the form of
\begin{equation}
    \pi(\theta | M) = \frac{1}{\sqrt{(2 \pi)^n \det G^{\alpha \beta}}} \exp\left(-\frac{1}{2} (\theta - \bar{\theta})^{\alpha} G_{\alpha \beta} (\theta - \bar{\theta})^{\beta}\right)
\end{equation}
come in especially handy as an analytical solution for the evidence exists. $G_{\alpha \beta}$ is the inverse covariance matrix and $\bar{\theta}$ the mean of the prior. Calculating the evidence given in Eq.~(\ref{eqn_bayes_evidence}) and taking the logarithm results into
\begin{equation}
    \begin{split}
        \ln p(y|M) =& - \ln \mathcal{N}_{\likeli} - \frac{n}{2} \ln 2 \pi + \frac{1}{2} \ln \det G_{\alpha \beta} \\
         &- \frac{1}{2} y^i C_{ij} y^j - \frac{1}{2} G_{\alpha \beta} \bar{\theta}^\alpha \bar{\theta}^\beta \\
         &+ \frac{n}{2} \ln (2 \pi) - \frac{1}{2} \ln \det (F_{\alpha \beta} + G_{\alpha \beta}) \\
         &+ \frac{1}{2} (F+G)^{-1 \hspace{0.05cm} \alpha \beta} (Q_{\alpha} + G_{\alpha \gamma} \bar{\theta}^\gamma) (Q_{\beta} + G_{\beta \nu} \bar{\theta}^\nu)\,.
    \end{split}
\end{equation}
As the normalisation of the likelihood as well as the product $y^i C_{ij} y^j$ are the same for all models, one can neglect them during model selection which enhances numerical stability.
\begin{figure}[t]
    \centering
    \includegraphics[width=.48\textwidth]{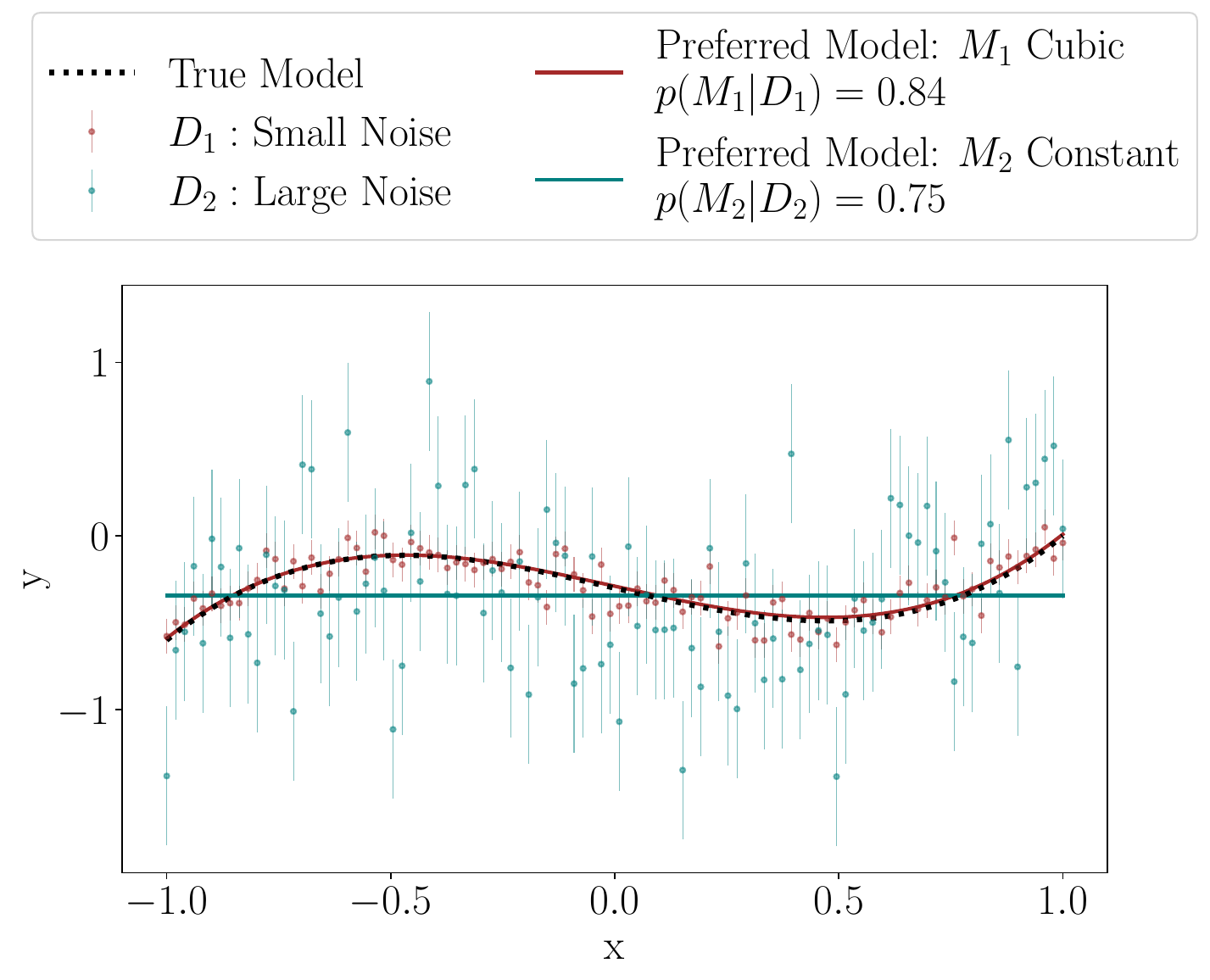}
    \caption{Two data sets are produced from the same model with small (brown) and large (teal) noise. The true model is of the form $y=\theta_0 + \theta_1 x + \theta_3 x^3$. The lines in the respective colours show the best-fit model obtained from the model posterior. For small noise, the best fit and truth coincide with a probability of 0.84. For large noise, a constant model is the most likely, yet with a probability of 0.75, as the effect from higher order terms is washed away by the noise.}
    \label{fig:toy_model_bestfit}
\end{figure}
Fig. \ref{fig:toy_model_bestfit} shows an example of generated polynomial data with two different noise levels and the least squared best fit of the found model. It shows the intuitive approach to model selection, where for large noise a simpler model, here a constant, is chosen over a more complex explanation.

\begin{figure*}[t!]
    \centering
    \includegraphics[width=\textwidth]{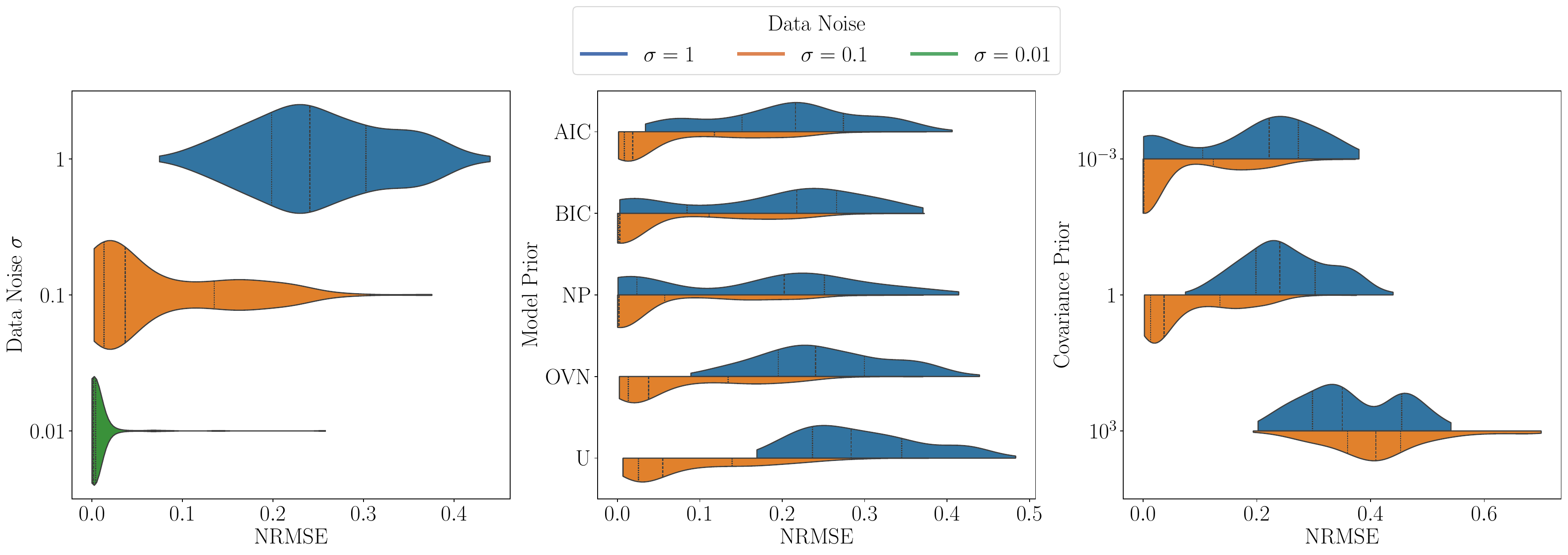}
    \caption{Toy model evaluation: (Left) Comparison of different noise levels in generated data - the less noise, the better the model selection. (Centre) Variety of model priors compared - stronger penalisation of more complex models performs better. (Right) For a Gaussian parameter prior different covariances were tested - a more narrow prior for suitable expectation value results into lower NRMSE (Normalised Root Mean Square Error).
    The drawn lines are the quantiles of the corresponding data set with the line in the centre being the median.}
    \label{fig:toy_model_results}
\end{figure*}

To get an idea of how the parameter and model priors influence the results, we generated three sets of toy data with different noise levels ($\sigma = [1, 0.1, 0.01]$). For each toy data model, we chose at random with a uniform probability whether a monomial is present or not. To make sure that we have different polynomial degrees in our data, i.e. low as well as high ones, we increased the allowed maximal polynomial degree every 40 toy datas resulting in 200 data sets per noise level. This means that our toy data includes polynomials of degree five. The results are shown in Fig.~\ref{fig:toy_model_results}. We use the Normalised Root Mean Square Error (NRMSE) as a measure for how well the preferred model fits the ground truth (gt). The NRMSE is defined as
\begin{equation}
    \text{NRMSE} (\kappa_{\text{gt}}, \kappa_{\text{fit}}) = \frac{\sqrt{\sum_\alpha (\kappa_{\text{gt}}^\alpha - \kappa_{\text{fit}}^\alpha )^2}}{\sqrt{\sum_\alpha (\kappa_{\text{gt}}^\alpha)^2}} \,.
\end{equation}
The left plot of Fig.~\ref{fig:toy_model_results} confirms, that less noise in the data resuls into a better model selction. Eventhough the NRMSE values are quite low, the model selection performs significantly worse for large noise as less information is extractable from the data.

For medium and large noise in the data, Fig.~\ref{fig:toy_model_results} centre compares the model selection results for different model priors. Considering medium noise (orange) the difference in NRMSE caused by the model priors is very minor and larger for bigger noise (blue). In both cases, the one over (OVN) or the uniform prior (U) perform worst. It is noteworthy that the normalisable prior (NP) is not only favourable for being normalisable, but also for performing best on our toy data. The good performance of the AIC, BIC and NP model priors can be explained by their strong penalisation of more complex models.

Lastly, while always using a Gaussian parameter prior, we vary the expectation value and the scaling of the prior covariance matrix. As one expects, the results are better the closer the expectation value is to the region of the data. The results for different prior covariances are shown in the right plot in Fig.~\ref{fig:toy_model_results}. While the difference in NRMSE is rather small for smaller covariances, a large prior covariance causes significantly worse results. This can be explained by the fact that the prior covariance now dominates the model evidence and thus the influence of the likelihood in terms of the Fisher information is minimised. Figuratively speaking, the model selection cannot take the data into account anymore and is mostly driven by the prior.

% ---  --- %
\section{Application to cosmology}\label{sec:supernova}

\subsection{Polynomial models for the dark energy equation of state}
The dynamics of the class of Friedmann-Lema{\^i}tre-Robertson-Walker spacetimes is fully characterised by the equation of state $w$ of the various cosmological fluids as well as their densities. As a consequence of the cosmological principle, they can only depend on time, and in many cases they reflect some property of the substance in question, such as $w = 0$ for matter and $w = +1/3$ for radiation. Evolving equations of state $w(a)$ are discussed in the context of dark energy cosmologies, where we choose to work with a polynomial
\begin{equation}
 w(a) = w_0 + w_1 (1-a) + \frac{w_2}{2} (1-a)^2 + \cdots = \sum_{j=0}^d \frac{w_j}{j!} (1-a)^j
\label{eq:des_definition}
\end{equation}
of order $d$, where the expansion in $1-a$ was chosen reminiscent of the CPL-parameterisation \cite{chevallier_accelerating_2001, linder_exploring_2003, 2009Caldwell, WeinbergCosmologicalConstant} that enforces $w(a) = w_0$ today. Considering only the first two terms of Eq.~(\ref{eq:des_definition}) results in the CPL-model, tested by many cosmological observations \cite{desicollaboration2024desi, pantheon+data}. We constrain the dark energy equation of state with type Ia supernovae from the Pantheon+ data set \cite{pantheon+constraints}, including 1590 supernovae in the redshift range $z=0.01\,\ldots\,2.25$.

For a given Hubble function we can calculate the distance modulus as
\begin{equation}
    \mu = m - M = 5 \log_{10}(d_L(a,\theta))+10\,,
\end{equation}
with the apparent magnitude $m$ and the absolute magnitude $M$ as well as
\begin{equation}
    d_L(a,\theta)= \frac{c}{a} \int_a^1\dd a^\prime \frac{1}{H(a^\prime,\theta)}\,,
    \label{eq:lumi_distance}
\end{equation}
where $a$ is the scale factor and $\theta$ is the collection of all relevant cosmological parameters. The luminosity distance in Eq.~(\ref{eq:lumi_distance}) can be solved as a numerical integral or, equivalently, as a differential equation. In the considered model, the Hubble function is given by
\begin{equation}
    \frac{H(a,\theta)^2}{H_0^2} = \frac{\Omega_m}{a^3}+(1-\Omega_m)\exp\left[-3\int_1^a\dd a^\prime \frac{1+w(a^\prime)}{a^\prime}\right]\,.
    \label{eq:hubble_function}
\end{equation}
We fix the Hubble parameter at ${H_0=70\,\text{km/s/Mpc}}$ and keep $\Omega_m$ and $M$ as nuisance parameters. The integral in the exponent of Eq.~(\ref{eq:hubble_function}) can be solved analytically for our polynomial parameterisation, defined in Eq.~(\ref{eq:des_definition}),
\begin{align}
    I&=\int_{1}^{a}\dd a^\prime \frac{1}{a^\prime}\left(1 + \sum_{j=0}^d \frac{w_j}{j!} (1-a^\prime)^j\right) \nonumber \\
    &= \ln(a)\left(1+\sum_{j=0}^d \frac{w_j}{j!}\right) + \sum_{j=1}^d w_j \sum_{l=1}^j \frac{(-1)^l}{l}\frac{a^l-1}{l!(j-l)!}\,,
\end{align}
at any order $d$.

% % ---  --- %
\subsection{Evidence Calculation}
\label{sec:supernova_evidence}
% ---  --- %
\begin{figure*}[t!]
    \centering
    \includegraphics[width=\textwidth]{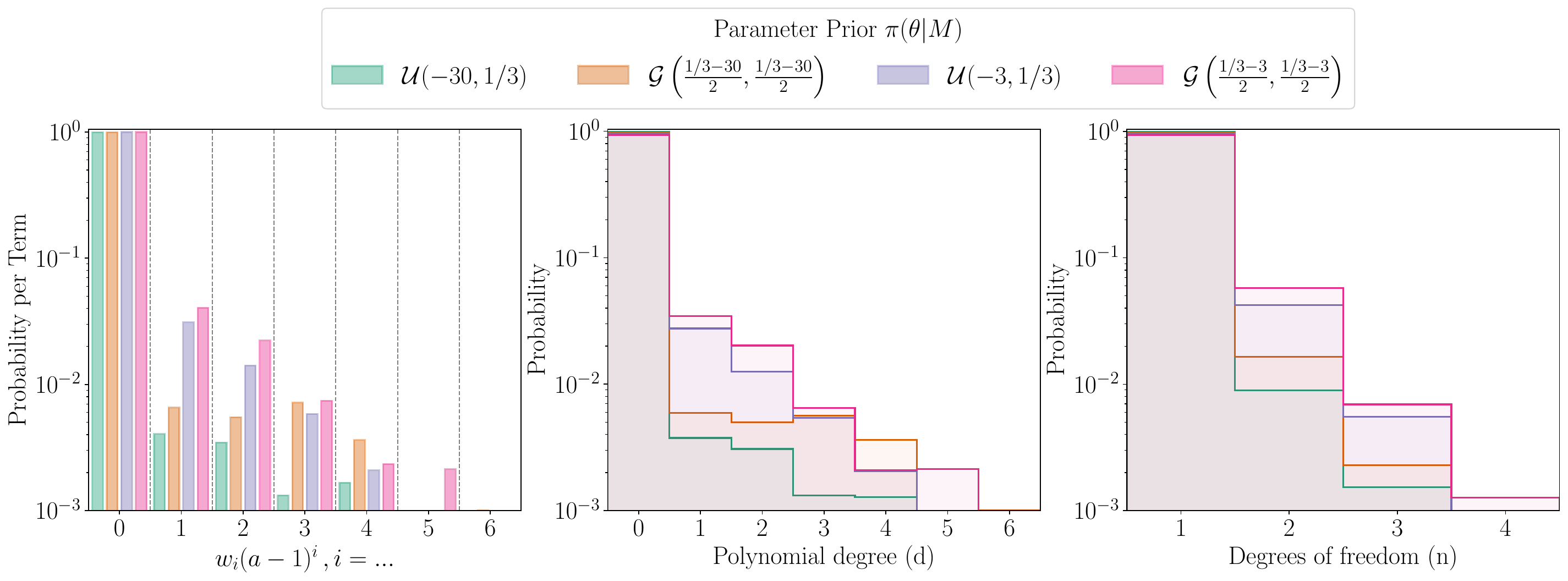}
    \caption{The three figures show the effect of four sensible prior choices for the model parameters on the model posterior. Values smaller than $10^{-3}$ are cut off. (Left) Each bar is the probability that the term of the polynomial is present in the model, independent of the others. (Centre) The marginalised posterior such that the histogram shows the probability distribution for the polynomial degree and (right) similarly for the degree of freedom. }
    \label{fig:subernova_param_prior}
\end{figure*}
Calculating the evidence for a model requires the likelihood and the parameter prior as defined in Eq.~(\ref{eqn_bayes_evidence}). The likelihood, under the assumption of a Gaussian error process, is given by
\begin{equation}
    \ln \mathcal{L}(y|\theta) =-\frac{1}{2}(m(\theta)-m_\text{obs})^i C_{ij}(m(\theta)-m_\text{obs})^j+\text{const.}\,,
\end{equation}
with the inverse covariance matrix $C_{ij}$, given in \cite{pantheon+data} and the theoretical apparent magnitude $m(\theta)$ as well as the observed value $m_\text{obs}$. There might be a need for more complex distributions, as explored in \cite{lovick2023nongaussianlikelihoodstypeia}. The parameters $\theta$ are the coefficients of the polynomial, $\Omega_m$ and $M$. For the nuisance parameters, we choose flat priors such that $\Omega_m\in[0,1]$ and $M\in[-22, -17]$. Priors for $w_i$ require more consideration because less is known about their value (in combination) and since we compare models with a different number of $w_i$ volume effects pertaining to the parameter spaces matter. 

We consider every term of Eq.~(\ref{eq:des_definition}) separately because we do not expect that a combination of unreasonably high values should cancel each other out in such a way that the polynomial is in the desired range. However, it is not entirely clear what range that would be. We investigate four parameter prior choices and their effects on the resulting posterior. One motivation for an upper limit is to require positive Ricci curvature, yielding a limit of $+1/3$. Yet, this is not the only option and many more possibilities are available. For the lower limit, we test multiple options and show their effect. The first one is $w_i\in[-3,1/3]$ inspired by the choice in \cite{desicollaboration2024desi}. The second one is supposed to be a lot wider with $w_i\in[-30,1/3]$, clearly allowing phantom cosmologies with diverging scale factors after finite times. Additionally, we check a Gaussian prior where the limits from the uniform prior are now at the inflection points. This should allow the likelihood to dominate over the prior if the maximum is outside the main region favoured by the data itself.

Evaluating the evidence requires numerical tools. One option is to use MultiNest \cite{Feroz_2009} with its Python implementation PyMultiNest \cite{pymultinest}. The dimensionality of the parameter set differs a lot depending on the considered model, it is at least three dimensional and was in practice up to 13 dimensional. Suitable settings for our precision requirements are 400 live points and an evidence tolerance of 0.5. We account for the numerical uncertainty by assuming that it is Gaussian and every time needed sampling from its distribution in the acceptance step. 

% % ---  --- %
\begin{figure*}[t!]
    \centering
    \includegraphics[width=\textwidth]{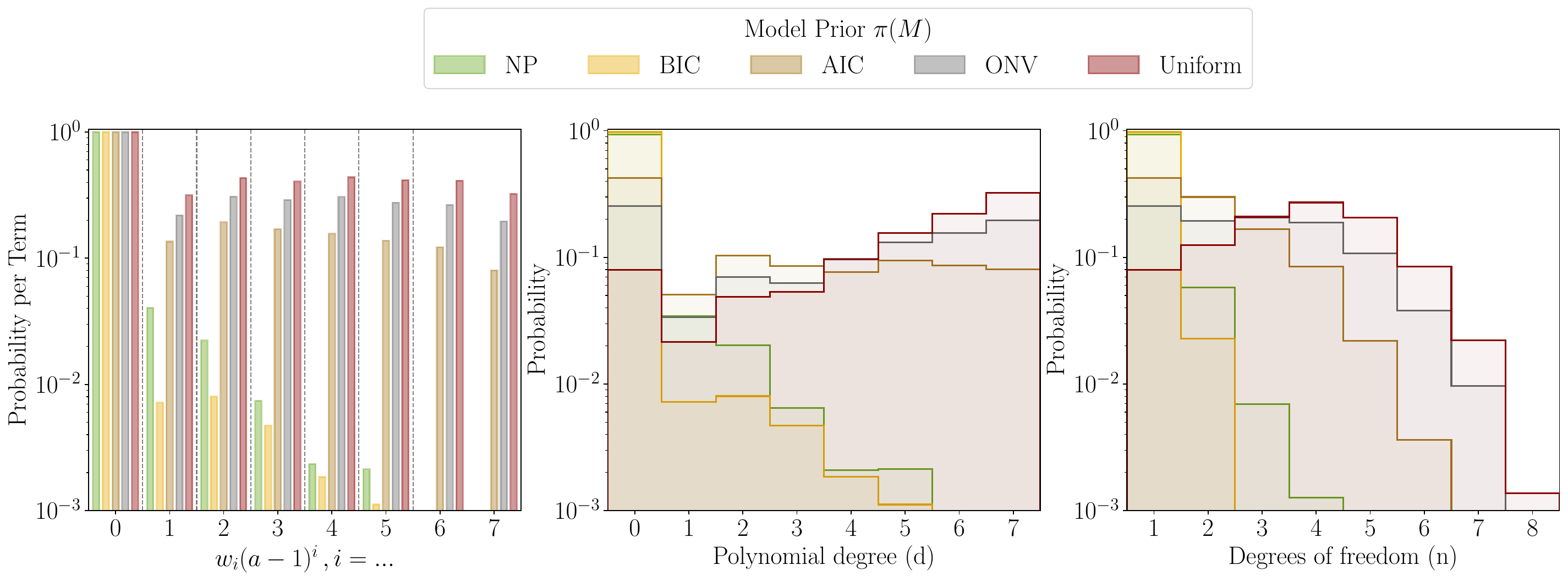}
    \caption{The three figures show the effect of sensible model prior choices on the model posterior. The prior formulas are given in Tab.~\ref{tab:IC_priors}. Values smaller than $10^{-3}$ are cut off. (Left) Each bar is the probability that the term of the polynomial is present in the model, independent of the others. (Centre) The marginalised posterior such that the histogram shows the probability distribution for the polynomial degree and (right) similarly for the degree of freedom. }
    \label{fig:subernova_model_prior}
\end{figure*}

\subsection{Posterior Model Probabilities}
Following the considerations of different model and parameter priors in Sect.~\ref{sec:toy_model} we investigate their effects in this application. To illustrate their effects, we will again use (i) the mean, see Eq.~(\ref{eq:mean_keyspace}), which is interpretable as the probability that one individual term is present in the polynomial, independent of all the others. (ii) is the marginalisation of the posterior over d.o.f, i.e. it is the probability of the model having a certain polynomial degree $d$. (iii) is the marginalisation over the polynomial degree, yielding the probability of having a certain number of parameters $n$ in the models, independent of their polynomial degree.
We start with the effects of the parameter priors, displayed in Fig.~\ref{fig:subernova_param_prior}. The four different choices are discussed in Sect.~\ref{sec:supernova_evidence} and are two uniform and two Gaussian priors, with different widths. The model prior is chosen as the normalisable one. We can see, that for all prior choices it is almost certain, that there is $w_0$ term and the distributions for polynomial degree and d.o.f peak strongly at zero and one, respectively. How fast the curve decays away from the maximum depends on the width and type of the prior. Narrower priors show a slower decay because they start to dominate over the likelihood if there is not enough information for the parameter set. This then results in a larger evidence, because we can approximate it by $p(y|M_i)\approx \delta \theta / \Delta \theta \, \hat{\mathcal{L}}(y|\theta,M_i)$ with $\delta \theta$ and $\Delta \theta$ the width of the likelihood and prior, respectively. The Gaussian prior is not as restrictive, which leads to a possibly higher best-fit likelihood value $\hat{\mathcal{L}}$ in the considered region. With the previous approximation, we can see that a Gaussian prior results in a higher evidence, if we have too many parameters and the likelihood is not as informative. However, the main findings are independent of the prior and for future results we use the narrow Gaussian prior.

Next, we study the different model priors described in Sect.~\ref{sec:prior_choice}. All priors, except the normalisable, need an upper boundary in the polynomial degree to be properly defined. Our choice is $d_\text{max}=7$, because this allows a vast model landscape, but is still feasible to cover. The different results are shown in Fig.~\ref{fig:subernova_model_prior}. While the main result, that the most likely model is $w_0$ and that there must be a constant term, is independent of the model prior, there are clear differences between them. Once the discrete Markov walk is in a region where the information in the data is not sufficient to properly constrain the parameters of the model, the evidence differs only marginally, as long as there is a constant term. Therefore, the upper boundary on the highest degree is not only necessary from a normalisability point of view but also from a practical one. It is most prominent in the uniform 'U', 'OVN' and 'AIC' priors, where the probability for the existence of the terms $w_i(a-1)^i,\, i\geq 1$ stays almost constant up to the boundary. The normalisable 'NP' prior punishes a high degree, independent of the number of parameters, hence, the expected drop off in probability. Similarly, the 'BIC' prior, does not explicitly discourage the chain from walking into regions with large $d$, but the large number of data points results in a very strong prior and it stays at the top of the model triangle. When we look at the marginalised probabilities in the centre and left plot of Fig.~\ref{fig:subernova_model_prior} it seems like the weaker priors induce a favour towards more degrees of freedom and a larger highest degree. However, these are common Bayesian marginalisation effects. There are many more models with these properties and summing over them leads to the increased probability. Yet, when considering the full two-dimensional posterior, the MAP model is always $w_0$.
\begin{wrapfigure}[22]{r}{0.5\textwidth}
    \vspace{-.55cm}
    \includegraphics[width=.5\textwidth]{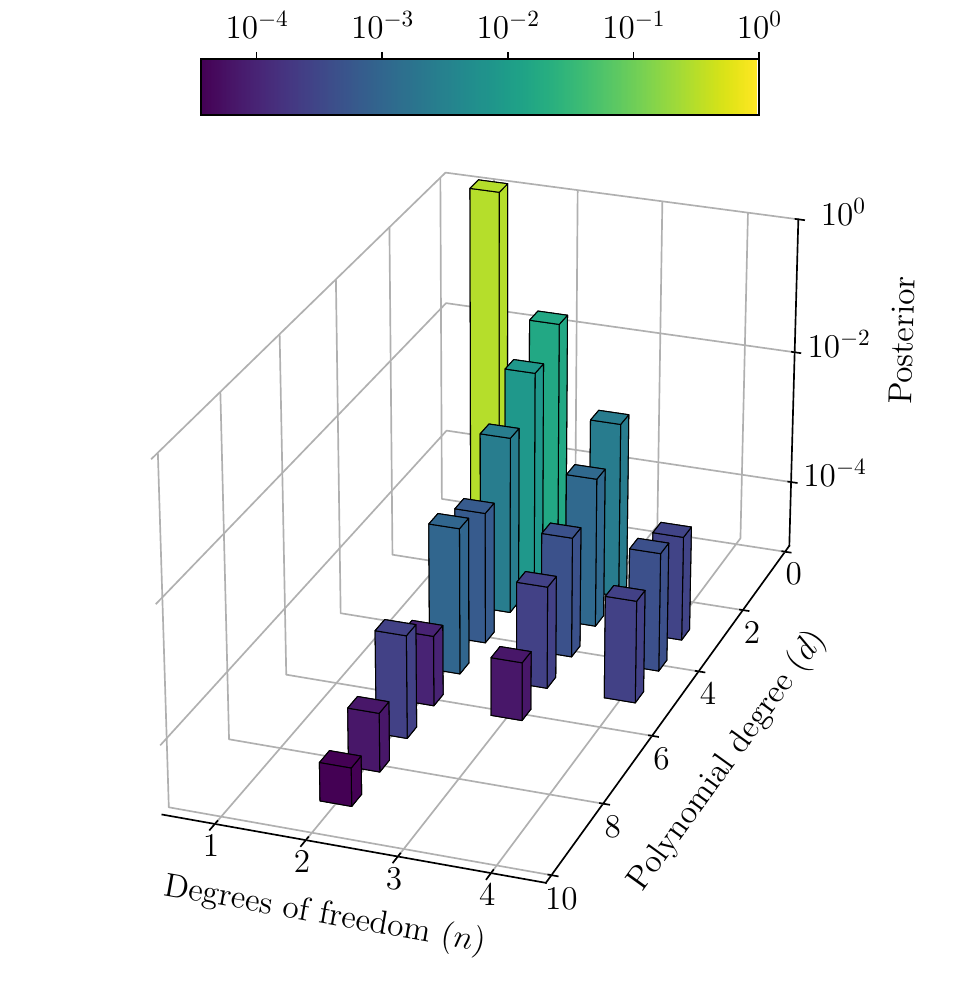}
    \caption{Model posterior for supernova data derived with a narrow Gaussian prior on the parameters and a normalisable prior on the models. The triangle outlay is the same as in Fig.~\ref{fig:model_space_polynoms}, where the probabilities for models with the same $n$ and $d$ are added. The height of the bar and its colour correspond to the posterior probability.}
    \label{fig:supernova_posterior}
\end{wrapfigure}\\

Finally, we take a look at the full model posterior. We think that a prior without sharp boundaries is to preferred over others with, to avoid surprise effects outside the boundaries of the prior. Therefore we choose the 'NP model prior for the final inference, combined with the narrow Gaussian on the parameters. We justify our choice for the parameter prior with the results from Sect.~\ref{sec:toy_model}, where an overly wide prior hindered correct model selection. The posterior $p(M_i|y)$ is shown in Fig.~\ref{fig:supernova_posterior}. The MAP model is $w_0$ with $p(w_0|y)=0.934$. The second best model is $w_0w_1$ with $p(w_0w_1|y)=0.034$. Additionally, we see that only a fraction of the model space has been explored. This is a clear advantage of the Markov walk, where one does not need to calculate the evidence for models in bad regions and therefore saves valuable computation resources, while still getting a properly normalised probability distribution. However, as with any MCMC technique, discrete or continuous, one has to be careful with multimodality. Yet, unlike in the continuous case, here one has more options to avoid it. It is possible to set up the model space in a different way, such that similiar models are closer, depending on the physics task. Additionaly, once the most probable models have been investigated it is computationaly very cheap to drastically increase the length of the chain, since all relevant evidences have been calculated and can be reused. 

% % --- section: summary --- %
\section{Conclusion}
\label{sec:conclusion}
Subject of this paper was the exploration of model spaces by means of Markovian random walks as an embodiment of Bayesian model selection. We addressed the problem from both the theoretical point of view, by bridging to the theory of partition functions, as well as from an application point of view, by constructing an algorithm for the evaluation of posterior model probabilities. As an application in cosmology, we computed posterior model probabilities for polynomial dark energy equations of state, as they are constrained by supernova data.

\begin{enumerate}[label=(\roman*)]

\item{The essence of Bayesian model selection is striking the balance between better fits provided by more complex models and the preference for simplicity. In the case of just two models, the Bayes factor $B$ is a criterion on which a decision can be based. In situations with more than two models, there is the need for generalisation of this pairwise model comparison to the exploration of an entire landscape of models, along with a proper evaluation of the posterior model probability $p(M|y)$ that takes the model prior probability $\pi(M)$ into account.}

\item{The random walk across the model space and the transition from model $i$ to model $j$ is governed by the ratio of the Bayesian evidences $p(y|M_j)/p(y|M_i)$. Rewriting this ratio as $\exp(-\Delta B)$ with $\Delta B = \ln(p(y|M_i)/p(y|M_j))$ recovers the Boltzmann factor $\exp(-\Delta B)$, which assumes a value of $\simeq 1/3$ for $\Delta B\simeq 1$. One might wonder if that is the numerical origin of the steps of 3 in Jeffreys' scale for interpreting evidence ratios and quoting them in terms of posterior odds \cite{Jeffreys:1939xee, Kass:1995loi, nesseris_is_2013}.}

\item{Following Occam's razor, the model prior should penalise complex models. By applying the maximum a posteriori and approximating the likelihood in zero order it was shown that the Akaike and Bayesian information criteria can be related to corresponding model priors. Additionally, a normalisable prior was introduced that also penalises high polynomial degrees.}

\item{The sampling properties of the Markovian walk are analytically described by means of canonical partition functions $Y[\zeta]$. They are constructed from the model evidence $p(y)$ as a weighted sum over the Bayesian evidences $p(y|M_i)$ and model priors $\pi(M_i)$. In their construction they reflect analogies to the canonical partition $Z[\beta]$ composed of the likelihood $\mathcal{L}(y|\theta,M)$ and the prior $\pi(\theta|M)$, which in turn define the posterior distribution $p(\theta|y,M)$. Consequently, $Y[\zeta]$ allows for a Bayesian inversion to yield the model posterior probability $p(M|y) = p(y|M)\pi(M)/p(y)$.}

\item{Constructing a partition function $Y[\zeta]$ by enhancing the expression for $p(y)$ with an inverse temperature $\zeta$ allows accessing quantities like the Shannon entropy of the posterior distribution $p(M|y)$: It results from differentiation of $\ln Y/\zeta$ with respect to $\zeta$, taken at $\zeta = 1$. Related partitions, with a modified weighting of the model prior with the exponent $\zeta + 1/\zeta$, give rise to the Kullback-Leibler divergence between $p(M|y)$ and $\pi(M)$, generalising the surprise statistic to model selection.}

\item{For the specific physics example, namely dark energy equation of state constraints from supernovae, the posterior probability singles out constant equations of state with numerical values close to $w=-1$, substantiating the case of a cosmological constant $\Lambda$. The actual posterior distribution of the simplest model with a constant polynomial is $w_0 = -0.94^{+0.17}_{-0.14}$, marginalised over $\Omega_m$ and $M$. This statement is not strongly dependent on the priors that we investigated, and both the properties of the equation of state polynomials as being constant and as having only a single term are favoured by the data.}

\item{Choosing a suitable model prior $\pi(M_i)$ helps to improve the model selection, but is not absolutely necessary as our method turns out to be quite robust against the choice of the model prior. Things are a bit different when it comes the parameter priors $\pi(\theta|M_i)$. Here, one must be careful in choosing a sensible prior region to avoid the evidence being prior dominated.}

\end{enumerate}

In conclusion, we formulate the idea that in the prior choice of $\pi(M_i)$ it should be possible to follow the principle of maximised entropy similarly as in the choice of $\pi(\theta|M_i)$ to keep the model selection unbiased.

\acknowledgments
\paragraph{Funding information}
This work was supported by Vector Stiftung. We acknowledge the usage of the AI-clusters {\em Tom} and {\em Jerry} funded by the Field of Focus 2 of Heidelberg University.

\paragraph{Data availability}
Our Python implementation of the code that explores the dark energy model landscape on the basis of supernova data, or polynomial toy data is available on \href{https://github.com/cosmostatistics/ebms_mcmc}{GitHub}.
% --- bibliography --- %

\bibliographystyle{JHEP}
\bibliography{references}
\end{document}